\documentclass[twocolumn,preprintnumbers,amssymb,amsmath,9pt]{revtex4}[10pt]
\usepackage{graphicx}% Include figure files
\usepackage{dcolumn}% Align table columns on decimal point
\usepackage{bm}% bold math
\usepackage{amssymb}
\usepackage{epsfig}
\newcommand{\be}{\begin{equation}}
\newcommand{\ee}{\end{equation}}
\newcommand\beq{\begin{eqnarray}} 
\newcommand\eeq{\end{eqnarray}}

\newcommand{\lya}{{Ly$\alpha$}}

\begin{document}

\title{White Noise from Dark Matter: 21 cm Observations of Early Baryon Collapse}

\author{Kathryn M. Zurek}
\affiliation{Department of Physics, University of Wisconsin, Madison, WI 53706}
\author{Craig J. Hogan}
\affiliation{Departments of Physics and Astronomy, University of Washington, Seattle, WA 98195}
\preprint{}
\begin{abstract}
In concordance cosmology, dark matter density perturbations  generated by inflation lead to nonlinear, virialized  minihalos, into which baryons collapse at redshift $z \sim 20$. We survey here novel baryon evolution produced by a modification of the power spectrum  from white noise density perturbations at scales below $k \sim 10 h \mbox{ Mpc}^{-1}$ (the smallest scales currently measured with the Lyman-$\alpha$ forest).  Exotic dark matter dynamics, such as would arise from scalar dark matter with a late phase transition (similar to an axion, but with lower mass), create such an amplification of small scale power.  The dark matter produced in such a phase transition collapses into minihalos, with a size given by the dark matter mass within the horizon at the phase transition.  If the mass of the initial minihalos is larger than $\sim 10^{-3} M_\odot$, the modified power spectrum is found to cause  widespread baryon collapse earlier than standard $\Lambda$CDM, leading  to earlier gas heating.  It also results in higher spin temperature of the baryons  in the 21 cm line relative to $\Lambda$CDM at redshifts  $z > 20$ if the mass of the minihalo is larger than $1 M_\odot$. It is estimated that experiments probing 21 cm radiation at high redshift will contribute a significant constraint on dark matter models of this type for initial minihalos larger than $\sim 10 M_\odot$.  Early experiments reaching to $z\approx 15$ will constrain minihalos down to $\sim 10^3 M_\odot$. %More massive lumps lead to higher gas temperatures, allowing cooling and collapse into stars at  higher redshift; for initial lumps  $\gtrsim 10^3 M_\odot$, baryon collapse and cooling begin shortly after recombination. 

%, which accompany formation of certain types of very low mass scalar dark matter. %Such a power spectrum injects large density perturbations on small scales, while leaving the inflationary spectrum of perturbations unmodified on large scales.  
%An early  phase transition  produces  dark matter minihalos, with a size given by the inverse particle mass, that collapse at matter/radiation equality.  

\end{abstract}

\maketitle

\section{Introduction}

The concordance cosmological model  (Cold Dark Matter with cosmological constant, $\Lambda$CDM) explains cosmological structural data over  four orders of magnitude in linear scale. Cosmic background anisotropy \cite{Spergel:2006hy}, galaxy surveys \cite{Adelman-McCarthy:2005se,Cole:2005sx},  and Lyman-$\alpha$ forest statistics \cite{2006ApJS..163...80M} all accord with a model of  non-interacting, non-relativistic matter, whose density perturbations are seeded by inflation and are nearly scale free.  As far as the present cosmological data on these scales are concerned, the particle properties of the dark matter are irrelevant, provided that the dark matter candidate is sufficiently cold and collisionless; the effects of the dark matter are set not by the internal properties of the dark matter itself (e.g. mass and interactions), but by the perturbations imprinted at very early times through inflation \cite{Loeb:2005pm}.  The modifications to the inflationary perturbation spectrum by the leading dark matter candidates, the Weakly Interacting Massive Particle (WIMP) and the axion, are so small that their effect on structure formation will be very difficult to observe.  For the WIMP, these modifications arise only at earth mass scales \cite{Diemand:2005vz}, while for the axion, the modifications occur at an even smaller scale of $10^{-12} M_\odot$ \cite{Kolb:1995bu}.

%The dark matter problem remains one of science's great unsolved mysteries.  Whatever the dark matter's presently unknown, microscopic internal properties, we do know, however, that it is observably well described by cold, collisionless matter, as has been well-confirmed through measurements of the power spectrum through low-redshift galaxy surveys \cite{Adelman-McCarthy:2005se,Cole:2005sx}, Lyman-$\alpha$ forest measurements \cite{2006ApJS..163...80M}, and high-redshift CMB measurements \cite{Spergel:2006hy}.  The effects of the dark matter on the observables of these experiments are set not by the internal properties of the dark matter itself (e.g. mass and interactions), but by the gravitational potential imprinted at very early times through inflation. 

This idealized cold and collisionless model of dark matter, however, may not apply on small scales.  The physics of dark matter may be consistent with all current data but lead to new and measurable effects on smaller scales, yielding signatures which are unique to each candidate. For example, there has been much work done to explain discrepancies between the predictions of $\Lambda$CDM and the observed number of dwarf satellites and their central densities by modifying the small scale dynamics of dark matter.  These include adding dark matter interactions \cite{Spergel:1999mh} or annihilations \cite{Kaplinghat:2000vt} with cross-sections roughly the size of hadronic cross-sections.  Other models modify the small scale power (on \lya\ scales of $k \simeq 10 h/\mbox{Mpc}$) by changing the free-streaming scale through late warming of the dark matter, as in certain supersymmetric models with a late decay of the next-to-lightest supersymmetric partner \cite{Feng:2003uy,Strigari:2006jf}, or through primordially warm keV mass sterile neutrino dark matter \cite{Dodelson:1993je,Shi:1998xs,Abazajian:2001nj}.

%All dark matter candidates do, in fact, modify the inflationary power spectrum in a way that is unique to each candidate. The most popular candidates, however, generally only give rise to modifications to the inflationary spectrum at scales well below the reach of current techniques, which can measure to $10^8 M_\odot$ through the Lyman-alpha forest.  Weakly Interacting Massive Particles (WIMPs), for example, give rise to modifications in the inflationary spectrum at the WIMP free-streaming scale, typically $\sim 10^{-9} M_\odot$.  QCD axions hardly give rise to a more notable signature, deviating only from the flat inflationary spectrum at scales $\sim 10^{-13} M_\odot$.  

Here we are interested in a different generic modification to small scale behavior   introduced by new dark matter physics: that associated with added fluctuation power on small scales, in particular additional  power with a white noise spectrum.  As discussed below, this added noise is a generic byproduct of the process that makes scalar dark matter during a phase transition. Since the spectrum is known the effects are characterized by one number, the amplitude of the added noise.

  Without loss of generality for the scales considered here, we choose to characterize this extra fluctuation power as   shot noise arising from discrete primordial lumps of dark matter, such as black holes or minihalos.  When the universe becomes matter dominated they  give rise to an additional power spectrum:
\be
P_{wn} = \frac{1}{n_{H}}=\frac{M_{H}}{\rho_{DM}},
\label{white-noise}
\ee
where $M_H$ denotes the mass of the lumps (be they halos or holes).
Note that such a power spectrum is wavenumber independent so that at sufficiently large $k$ it dominates over the inflationary spectrum, which drops as $k^{-3}$.

Primordial black hole dark matter is  itself highly constrained by astrophysical effects other than their effect on the power spectrum;   in particular, strong gravitational lensing experiments now rule out primordial black holes with $M_{H} \gtrsim 10^{-7} M_\odot$ (with the possible exception of a small window around $100 M_\odot$). On the other hand for scalar  dark matter the primordial minihalos are not particularly compact, but have a density determined by the first self-gravitating collapse,  at around the epoch of matter domination; thus the dark matter minihalos are not always of negligible size but instead  are gravitationally bound diffuse systems, hence the name ``minihalos''. In this case   the lensing properties of the minihalos (termed ``scalar miniclusters'' as studied in \cite{Zurek:2006sy}) depend on the distance, mass and mass profile; current  bounds from  lensing, and from the  Ly$\alpha$ forest power spectrum measurements, allow $M_{H}$ as large as $\approx 4 \times 10^3 M_\odot$ \cite{Afshordi:2003zb,Zurek:2006sy}.  

Here we survey  the  effects of primordial white noise on early baryon evolution,  such as early gas collapse, heating, and star formation, and their possibly observable effects in forthcoming experiments \cite{Morales:2004ca,Morales:2005qk}  designed to directly  measure signals from hydrogen hyperfine transitions  at  the epoch of cosmic reionization and higher redshifts.   The standard  $\Lambda$CDM models of these events \cite{Barkana:2000fd,Furlanetto:2006jb,Fan:2006dp} show a significant range of possible behavior for $z<20$ after stars begin to form, since observables depend significantly on the output  of stellar populations.  We do not present such  detailed models here, but  estimate the range of $M_H$ for which various departures from standard cosmology are expected to be appreciable and observable at higher redshifts  before stellar activity becomes significant in $\Lambda$CDM.  

These are our main conclusions about the effects of the added white noise, depending on the mass $M_H$ of the initial seed minihalos (which in turn depends on the scalar particle mass):
\begin{itemize}
\item
Minihalos with  $M_H$ below about $10^{-3} M_\odot$ cause essentially no new effects. They create halos whose virial temperature is always too low to accrete baryons.
\item
Minihalos in the range  $ 10^{-3} M_\odot <M_H<1M_\odot$ add to inflationary power significantly, enough to influence baryon collapse at redshift $z<30$, but the added effects may be difficult to distinguish observationally from moderately rare ($\approx 2\sigma$)  inflationary perturbations.
\item
Minihalos  as small as $M_H \approx 1 M_\odot$ create halos whose virial temperature is high enough to accrete and heat baryons at $z\gtrsim 30$, well before these effects occur widely in $\Lambda$CDM (at $z< 20$).  This  modifies the standard predictions   for quantities affecting potential observables,  such as the  spin temperature of the gas in the range $20<z<50$, even though the  spatial distribution of gas on larger scales is little affected.   The departures have   clear signatures: the kinetic and spin temperatures of the gas  are higher than in standard cosmology and begin activity at much higher redshift. On the other hand the departure of spin temperature is rather small at $z\lesssim 50$, making detection difficult.

\item
Minihalos above $M_H \approx 10  M_\odot$  create halos whose virial temperature is high enough for $3\sigma$ fluctuations to result in  molecular and atomic cooling  at  $z>30$, allowing baryons to not only collapse, but also radiate, become self-gravitating and form stars much earlier than $\Lambda$CDM (which predicts the first stars at $z< 20$).   In this regime, experiments reaching beyond $z\approx 20$ should see significant differences from $\Lambda$CDM predictions.
Indeed, for lumps at the current limit $M_H \approx 10^3 M_\odot$, collapse and star formation are expected already at the epoch of recombination.  Star formation can  have observable effects via infrared emission and  enrichment.  In addition,  the spin temperature perturbation is many times  larger than the standard prediction at redshift $z>20$,  and appears in emission rather than absorption. 
\end{itemize}

We conclude that the future generations of high redshift experiments will likely provide new constraints on dark matter behavior, and may constrain initial  minihalo mass  as small as $M_H\approx 10 M_\odot$.  %There are other cadidates, however, that do give rise to more significant modifications of the power spectrum on small scales.  Warm sterile neutrino dark matter, for example, results in  modifications to the power spectrum on scales just below the Lyman-alpha limit.  By construction, the sterile neutrino mass is chosen such that it damps the power spectrum sufficiently on those scales to solve the dwarf satellite problem. More generally, any dark matter candidate that decoupled while being relativistic so that free-streaming erased small-scale power produces similar effects.   

%  We require that this not occur until $k > k_{\mbox{Ly}\alpha}$, the smallest scale measurable by the Ly$\alpha$ forest. This results in a mass limit $M_{pbh} \lesssim 3 \times 10^3 M_\odot$.

\section{Scalar Dark Matter and Extra White Noise Power}
 Generically, any  
 second order phase transition in the early universe, where a scalar field of very low mass evolves from fixed vev to oscillation about the potential minimum, generates density perturbations on the scale which is the horizon size at the phase transition, $d_H(T_{trans})$.  The density fluctuations which arise from such a phase transition have two sources: classical or quantum fluctuations of the field. Consider a complex scalar field, $\phi = r e^{i a/f}$; classical fluctuations come from the angular mode, $a$, and quantum fluctuations from the radial mode, $r$.  

The classical fluctuations arise from spatial variations in the vev of $a$ on scales larger than the horizon; on smaller scales, spatial gradients will align the field locally (the Kibble mechanism). When $H \sim m_a$, the field begins to oscillate around the true minimum at $\langle a \rangle = 0$, and dark matter will appear with density $\rho = \frac{1}{2} m_a^2 \langle a \rangle_i^2$.  Since $\langle a \rangle$ has spatial fluctuations between $-\pi$ and $\pi$, the local dark matter density will have $O(1)$ density perturbations on the scale which is the horizon size at the time of the phase transition, $d_H(T_{trans})$.  The QCD axion, for example, is marked by such fluctuations when the Peccei-Quinn symmetry breaks after inflation;  the resulting structures, axion miniclusters,  have been studied extensively in \cite{Hogan:1988mp,Kolb:1993zz,Zurek:2006sy}.  (Standard axion miniclusters have masses $M_{h} \sim 10^{-12} M_\odot$, determined by the mass of axions within the  horizon at ~1 GeV when the axion mass becomes large enough to start oscillations and dark matter formation.)

The second means to generate density fluctuations, through quantum fluctuations, requires only slightly more explanation.  The size of the quantum fluctuations in the radial mode is set by its mass: $\langle \delta r \rangle \sim m_r$.  These fluctuations in $\langle r \rangle$ lead to fluctuations in the time (or, equivalently, the temperature) of the phase transition:  $\delta t_{trans} \sim \langle \delta r \rangle / \dot{\langle r \rangle}$, where $\dot{\langle r \rangle} \sim m_r \langle r \rangle$ is the evolution of the background field, so that $\delta t_{trans} \sim 1/m_r$.  The last observation is that the density fluctuations are proportional to the fluctuations in phase transition time (or equivalently, temperature)
\be
\frac{\delta \rho}{\rho} = 3 \frac{\delta T}{T} \sim H \delta t_{trans}
\ee
Since $\delta t_{trans} \sim 1/m_r$, $\delta \rho/\rho \sim H/m_r$.  For a phase transition occurring just as the particle mass enters the horizon, density fluctuations will again be $O(1)$.  The size of the density fluctuations can be suppressed if some finite temperature effects prevent the phase transition from occurring promptly, and instead $H/m_r < 1$. 

We have shown that the density fluctuations on the scale $d_H(T_{trans})$ are non-linear.  On smaller scales, density perturbations are damped since spatial gradients are able to align the field locally on  a light-crossing timescale.  On larger scales, the density perturbations are simply white noise, so that the total power spectrum is 
\be
P = \frac{M_H}{\rho_{DM}} e^{-(k d_H)^2/2}.
\label{wnpower}
\ee  
where $M_H$ is the dark matter mass contained in $d_H(T_{trans})$. Note again that since this power spectrum is $k$ independent up to the cut-off scale at $d_H$, the white noise spectrum will be insignificant at smaller $k$ where the inflationary perturbations are most important, but may dominate the power spectrum at smaller scales where the inflationary spectrum scales as $k^{-3}$.  A $10^3 M_\odot$ mass minihalo corresponds to a phase transition at 0.1 GeV or a $10^{-12} \mbox{ eV}$ mass scalar.

The white noise power spectrum of eqn.~\ref{wnpower} gives rise to novel, qualitatively different structure formation in the matter dominated era.  Due to the large density fluctuations on small scales, virialization occurs much earlier, generally as early as $T_{eq}$.  If the virialized potentials are deep enough, baryons   begin to collapse onto the objects very early,  altering the temperature and density of the gas; if further energy loss from cooling is efficient, the baryons can then cool to form stars.  %In addition, through early virialization, the kinetic temperature of the gas rises early, and may lead to distinct signatures in 21 cm tomography.  It is the purpose of this paper to survey these modified signatures from non-inflation-seeded structure formation.

\section{Structure formation with white noise power spectrum}

In $\Lambda$CDM structure formation, all density perturbations are initially linear.  After matter-radiation equality, they grow linearly with scale factor until the perturbations become $O(1)$.  Using the Press-Schechter analytic approach to structure formation, the small scale objects are incorporated into larger objects hierarchically as the perturbations on larger scales themselves become non-linear.  The first objects are typically formed at redshifts $z \sim 20$.

\begin{figure}
\includegraphics[width=8.5cm]{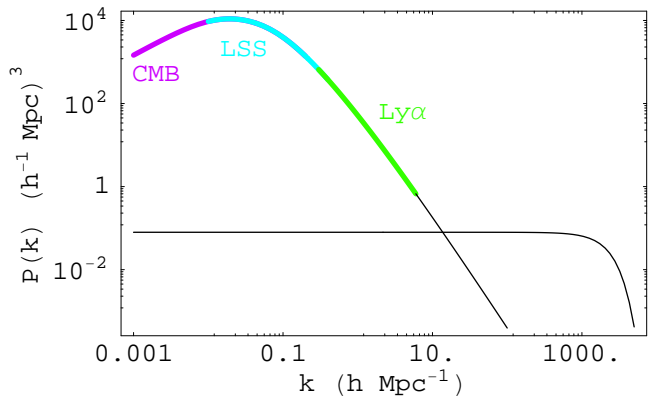}
\caption{Inflationary power spectrum and white noise power spectrum with $M_H = 10^3 M_\odot$.  CMB, LSS and \lya \ indicate the regions where the Cosmic Microwave Background, Large Scale Structure, and \lya  \ forest measurements constrain the power spectrum. }
\label{power}
\end{figure}

\begin{figure}
\includegraphics[width=8.5cm]{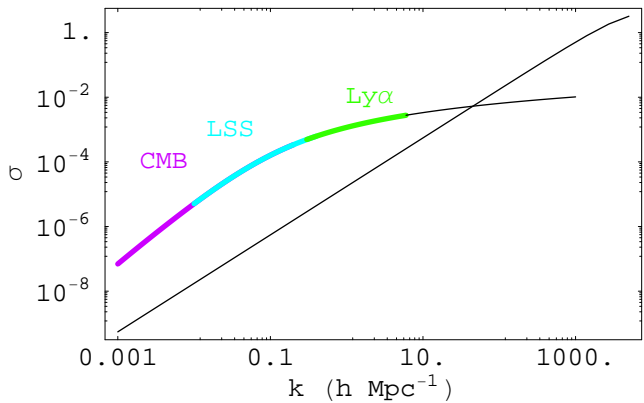}
\caption{Variance of inflationary perturbations and white noise power spectrum with $M_H = 10^3 M_\odot$.  CMB, LSS and \lya\ indicate the regions where the Cosmic Microwave Background, Large Scale Structure, and \lya\ forest measurements constrain the power spectrum. The added variance at small scales produces early baryon collapse and observable effects at high redshift that would have escaped detection by the other techniques.}
\label{flucs}
\end{figure}

\begin{figure}
\includegraphics[width=8.5cm]{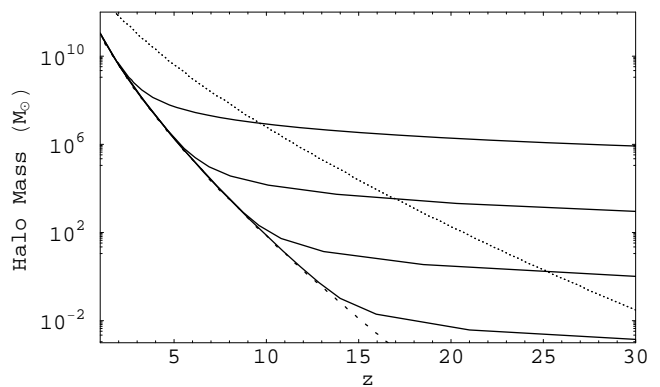}
\caption{Masses of typical collapsing halos as a function of redshift.  The dotted lines correspond to 1 and 2$\sigma$ fluctuations from an inflationary power spectrum.  The solid lines show the deviation from $\Lambda$CDM for the addition of white noise power with seed masses $M_H = 10^{-6} M_\odot, \mbox{ } 10^{-3} M_\odot, \mbox{ }  1 M_\odot, \mbox{ }  10^3 M_\odot$.}
\label{inflationary-power}
\end{figure}

\begin{figure}
\includegraphics[width=8.5cm]{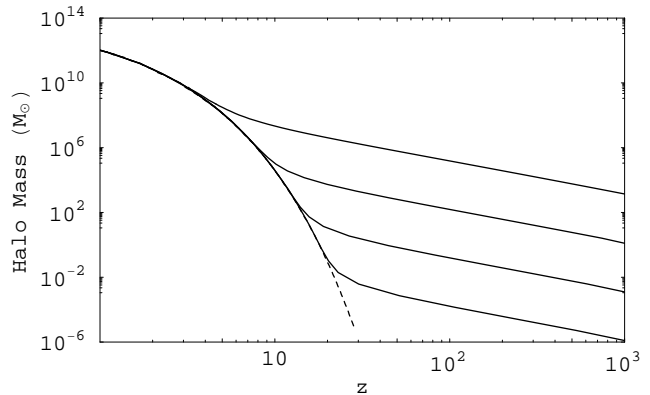}
\caption{Same as the previous figure, but extending to higher redshift.  The figure shows the larger deviations from $\Lambda$CDM at high redshifts.}
\label{white-noise-power}
\end{figure}

With the addition of white noise perturbations, we have a very different scenario. The initial power spectrum is characterized by non-linear dark matter fluctuations on the scale $d_H(T_{trans})$.  On larger scales,the perturbations decrease with  wavenumber as $k^{3/2}$.  On scales where the perturbation is initially non-linear, structure is immediately formed at matter-radiation equality.  On larger scales, the fluctuations grow linearly with redshift until they too become non-linear and collapse.  This is written as a fractional rms variation in density on a certain mass scale,
\be
\sigma(M,z)= \sigma(M)_i \frac{1+z_{eq}}{1+z}
\ee
where the variance is
\be
\sigma^2(M,z=0) = \int \frac{dk}{2\pi^2} P(k) \left[\frac{3 j_1(kR)}{kR}\right]^2,
\ee
$M = 4 \pi \rho_m R^3/3$ and $j_1(x) = (\sin x - x \cos x)/x^2$ is a particular choice of  window function for variance of mass within spheres.  Using the white noise power spectrum, the density fluctuations can be approximated as
\be
\sigma(M,z) \approx \sigma(M_H)_i \left(\frac{M}{M_H}\right)^{1/2} \frac{1+z_{eq}}{1+z}
\ee
where $\sigma_i$ is a number of order unity set by the details of the phase transition and the form of the cutoff chosen for the white noise power spectrum.  The initial power spectrum and density perturbations, $\sigma_i$ are shown in figs.~\ref{power} and \ref{flucs}.  Using the Press-Schechter approach, we expect an object of mass $M$ to collapse and virialize during the matter-dominated era at a redshift $z$ when $\sigma(M,z) \simeq 1.69$.  In fig.~\ref{inflationary-power}, we show the masses of typical collapsing halos both in $\Lambda$CDM (dotted lines corresponding to 1 and 2$\sigma$ fluctuations) and with added white noise (solid lines corresponding to seed masses $M_H = 10^{-6} M_\odot, \mbox{ } 10^{-3} M_\odot, \mbox{ }  1 M_\odot, \mbox{ }  10^3 M_\odot$).  In this last scenario, significant structure formation begins as early as $z_{eq}$.  The departures from $\Lambda$CDM become even more noticeable at higher redshifts, as shown in fig.~\ref{white-noise-power}.   We can see that at high redshifts, more massive halos virialize much earlier with added white noise power  

%In fig.~\ref{inflationary-power}, by contrast, we show structure formation from inflationary perturbations  alone (lower most line), and superimpose for comparison structure formation with white noise power spectrum added from fig.~\ref{white-noise-power}. .  This comparison is shown for ``typical'' (1-$\sigma$) halos; in standard $\Lambda$CDM, rare collapses start somewhat earlier (which is why the first stars make their appearance at $z\approx 20$ not $z\approx 10$) but with the white noise scenarios, widespread collapse of baryons can begin much earlier.
 
%where $\rho_{DM} (2\pi/k_H)^3 = M_H$, and

%Structure forms at $z_{form}$ when the perturbations have grown to be nonlinear, or more precisely, by Press-Schechter, when $\sigma(M,z_{form}) \sim 1.69$.  To show the non-standard structure formation, we plot in fig. the typical mass of collapsing halos as a function of redshift for four different seed masses, $M_H = 10^{-6} M_\odot, \mbox{ } 10^{-3} M_\odot, \mbox{ }  1 M_\odot, \mbox{ }  10^3 M_\odot$.  For comparison, we show structure formation from inflationary perturbations alone.

\begin{figure}
\includegraphics[width=8.5cm]{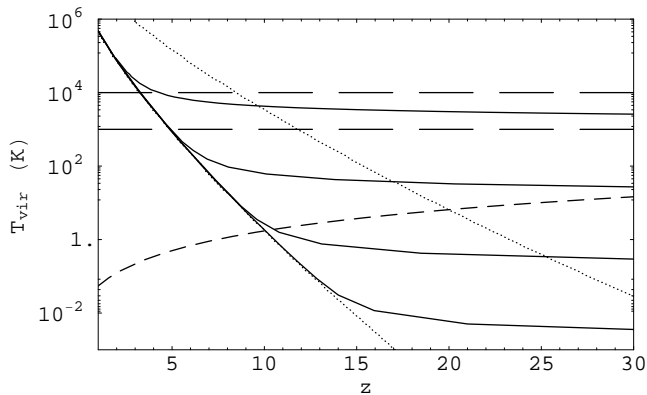}
\caption{Virial temperature (in Kelvin) of typical collapsing objects, as a function of redshift.  The dotted lines correspond to 1 and 2$\sigma$ fluctuations from an inflationary power spectrum.  The solid lines show the deviation from $\Lambda$CDM for the addition of white noise power with seed masses $M_H = 10^{-6} M_\odot, \mbox{ } 10^{-3} M_\odot, \mbox{ }  1 M_\odot, \mbox{ }  10^3 M_\odot$.  When the virial temperature rises above the temperature of the unperturbed baryons (shown by short dashed line), the baryons collapse onto the dark matter structure, leading to very early clumping and heating.  In addition, when the virial temperature is above $10^3$ and $10^4$ Kelvin (shown by long dashed lines), molecular and atomic cooling, respectively, may proceed, and the clumped baryons begin to form stars. Seeds above about $1 M_\odot$ cause widespread nonlinear gas collapse at $z>30$, significantly earlier than $\Lambda$CDM; those above
 $10^2M_\odot$ produce widespread  cooling and star formation at high redshift $z>30$; those above about $10M_\odot$ do so for 3$\sigma$ fluctuations.}
\label{inflationary-temp}
\end{figure}

\begin{figure}
\includegraphics[width=8.5cm]{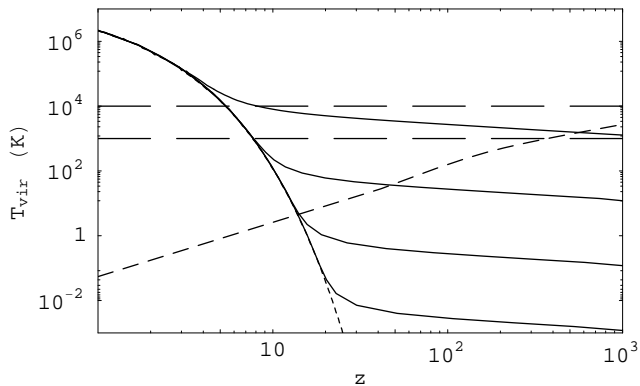}
\caption{Same as the previous figure, but extending to higher redshift.  The figure shows the larger deviations from $\Lambda$CDM at high redshifts}
\label{white-noise-temp}
\end{figure}

When the halo virial temperature of the collapsing halos rises above the temperature of the baryons, the baryons begin to accrete. The virial temperature for neutral primordial gas can be expressed in terms of the mass of the collapsing halo \cite{Barkana:2000fd}, 
\be
T_{vir}  \approx 4\times 10^4 \left(\frac{M}{10^8 h^{-1} M_\odot}\right)^{2/3}\left(\frac{\Omega_m}{\Omega_m^z}\right)^{1/3}\left(\frac{1+z}{10}\right) \mbox{K}
\ee
where $\Omega_m^z = \Omega_m(1+z)^3/(\Omega_r(1+z)^4+\Omega_m(1+z)^3+\Omega_\Lambda)$.  The virial temperature of typical collapsing halos as a function of redshift is shown in figs.~\ref{inflationary-temp} and \ref{white-noise-temp}, the analogs of figs.~\ref{inflationary-power} and \ref{white-noise-power}.  Because of the higher virial temperature associated with collapsing halos from the amplified power spectrum on small scales, baryons may collapse into halos at much higher redshifts.  The heating may be sufficient to start early star formation and reionization as well.

When the baryons have collapsed into the halos, they are then heated via shocks and adiabatic compression to the virial temperature of the halo.  If there is no further energy loss from cooling, the baryons then stabilize hydrostatically with gas pressure gradients balancing  the gravitational attraction of the combined halo and baryon mass. On the other hand if cooling is possible, the energy loss allows further collapse; eventually the baryons become self gravitating and the collapse becomes unstable until the formation of stars provides  heating to counteract the effect of cooling.  

At $z \gtrsim 200$, Compton scattering typically couples the baryons to the CMB photons, and uncollapsed baryons cool as $(1+z)^{-1}$.  When they decouple, uncollapsed baryons cool adiabatically, according to $(1+z)^{-2}$.  When $z \gtrsim 200$ and baryons have begun accreting onto the halos, radiative loss   occurs through Compton cooling; at any redshift, once the virial temperature reaches above $~10^3$ K and $~10^4$ K, molecular hydrogen and atomic cooling can occur, respectively.  Dashed lines have been added to figs.~\ref{white-noise-temp} and \ref{inflationary-temp} for the background baryon temperature, and the temperatures at which CMB, molecular, and atomic cooling become efficient.  When the virial temperature rises above these lines, widespread efficient star formation can occur.

These figures illustrate a few basic behavior thresholds. For $M_H$ less than about $10^{-3}M_\odot$, white noise  creates early-collapsing halos but these are below the baryon temperature so they cause only small amplitude perturbations in the gas.  There is no time at which baryon collapse occurs much differently from standard dark matter. For $M_H\approx 1M_\odot$, baryons collapse commonly at $z\approx 50$, much earlier than usual, and they are heated by adiabatic compression and shock heating, to temperatures well above the standard model, even without the action of stars. Presumably a closer study would reveal that these models also form some stars, in rare high density peaks able to cool,  and that this also happens much earlier than usual. For $M_H>10^2 M_\odot$ it becomes widespread (at the $1\sigma$ level) to form stars at $z>30$ above the molecular cooling threshold.
Finally, for $M_H\approx 10^{3}M_\odot$, the halos accrete gas already at recombination, at a temperature where cooling is efficient, so star formation is  widespread and the gas temperature well above the radiation temperature at all redshifts.

\section{Spin temperature heating and 21 cm tomography}

We do not consider in detail  heating and ionization from these first very early stars, or
the Wouthuysen-Field effect by which transitions induced by stellar \lya\  photons help to couple the spin temperature to the kinetic temperature of the gas.  Instead we aim to estimate simply the amount of white noise power needed to make an observable departure from the behavior of standard $\Lambda$CDM. It is a reasonable first approximation, even in the presence of stellar heating, to assume that the typical temperature of the bulk of diffuse gas is about the virial temperature of the dark matter halos; much hotter gas would be expelled and cool by expansion, whereas much colder gas would collapse and form stars, until the stellar feedback is sufficient to heat the gas and halt collapse. We know from surveys of baryons today (which are still mostly in diffuse hot ionized gas)  that most of the gas did not form stars in the end so such feedback must have operated rather efficiently.

A relatively clear signature of these effects may   appear in 21 cm transitions in gas at high redshifts. %The 21 cm hyperfine transition feature in neutral hydrogen is much weaker than the \lya\ line, so that its absorption saturates much less quickly through the intervening hydrogen.  As a result, it is a good probe over greater cosmological distances than the Lyman-alpha forest.
The observable effects on antenna temperature fluctuations scale with the spin temperature of the bulk of  relatively diffuse gas,  which in our scenario is significantly modified by the presence of the white noise.  To estimate the redshift and magnitude of departures from the standard model we adopt  here a simplified model of gas density,   kinetic and spin temperatures, adequate for identifying the thresholds of significant observable departures from $\Lambda$CDM.   A more complete account of the relevant physics, and more detailed models for $\Lambda$CDM,  can be found in \cite{Barkana:2000fd,Furlanetto:2006jb}.

The strength of the transition is characterized by the spin temperature, which relates the relative populations of the ground and excited states of the 21 cm transition through $n_1/n_0 = 3 e^{-T_*/T_S}$, where $T_* = 0.068 \mbox{ K}$.  This spin temperature is controlled by a balance of radiative and collisional transitions, so that the spin temperature can be related to the background radiation and kinetic temperatures, $T_\gamma$ and $T_K$:
\be
T_S = \frac{T_\gamma+yT_K}{1+y}.
\ee
$y$ depends on the radiative and collisional Einstein coefficients, so that $y \propto n_H$, the number density of hydrogen.  It is computed from the tables in \cite{Furlanetto:2006jb}.  $T_\gamma$ is the temperature of the radiation that is absorbed or stimulates 21 cm emission; this is  dominated by  CMB photons, $T_\gamma = T_{CMB}$. % We are interested in the 21 cm effects before the complicated effects of stellar feedback heat the gas and ionize it, so that the source of photons is the CMB, $T_\gamma = T_{CMB}$.  
If $T_S > T_{CMB}$, the resulting excess radiation brightness temperature through the cloud is given by 
\be
T_b = T_S (1-e^{-\tau})\simeq \tau T_S,
\ee
 where $\tau$ is the optical depth:
\be
\tau = \frac{3 A_{10} n_H}{16 \nu^2_0 T_S H(z)}.
\ee
Here $A_{10}= 2.85 \times 10^{-15} \mbox{ s}^{-1}$ is the spontaneous emission coefficient for the transition, and $\nu_0 = 1420.4 \mbox{ MHz}$ is the rest frame transition frequency.  Notice that when $T_S > T_{CMB}$ the transition is seen in emission, and the brightness temperature is independent of the spin temperature.  If, on the other hand, $T_S < T_{CMB}$, then the transition is seen in absorption, with brightness temperature \begin{eqnarray}
T _b &=& T_{CMB} e^{\tau} + T_S(1-e^{-\tau}) \nonumber \\
& \simeq & (T_S - T_{CMB}) \tau.
\end{eqnarray}  

\begin{figure}
\includegraphics[width=8.5cm]{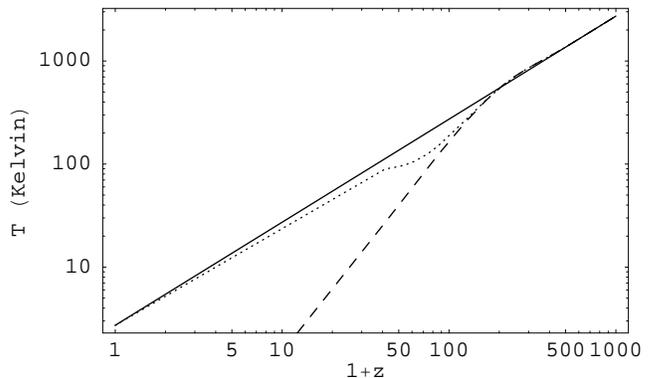}
\caption{Evolution of spin temperature (short dashed line) of uncollapsed baryons in (simplified) standard CDM due to coupling with CMB photons.  At intermediate redshifts, the spin temperature interpolates between the kinetic temperature of the baryons, $T_K$ (long dashed), and the CMB temperature $T_{CMB}$ (solid).  As a result, the 21-cm line is seen in absorption, $T_S < T_{CMB}$, at those redshifts.}
\label{spin-temp-cdm}
\end{figure}

In standard CDM, the baryons do not collapse onto nonlinear dark matter structures until $z \lesssim 20$, so that at higher redshift their temperature evolves uniformly with the background.  At $z \gtrsim 200$, Compton scattering couples the baryons to the photons, so that the temperature of the baryons is the same as the CMB temperature: $T_K \simeq T_{CMB}$.  The number density of hydrogen is high enough that $y \propto n_H \gg 1$ and $T_S \simeq T_K$.  Once the photons and baryons decouple, the uncollapsed baryons cool adiabatically, $T_K \sim (1+z)^2$.  The spin temperature continues to track $T_K$ until the number density of hydrogen becomes small enough that $y \lesssim 1$.  $T_S$ then again tracks $T_{CMB}$. As a result, at some intermediate redshifts, the spin temperate interpolates between the kinetic temperature and the photon temperature, and $T_S < T_{CMB}$.  We show this scenario in fig.~\ref{spin-temp-cdm}.  One may also in principle see density and temperature inhomogeneities in  the intervening hydrogen through absorption of CMB photons in the 21 cm line \cite{Loeb:2003ya}. 

%When early structure formation occurs, however, the kinetic temperature of the gas rises very early through virialization, so that the kinetic temperature of the gas does not simply cool adiabatically once the photons and baryons decouple.   

\begin{figure}
\includegraphics[width=8.5cm]{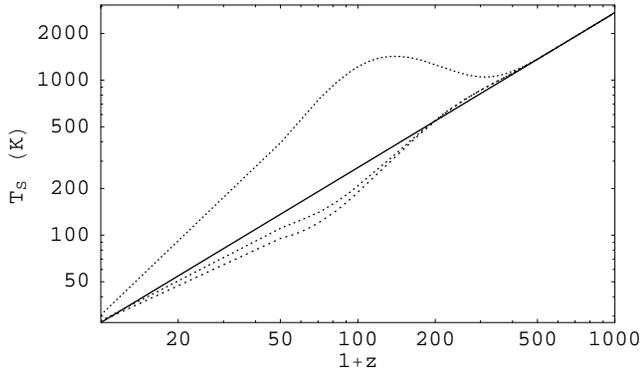}
\caption{Spin temperatures of newly formed objects, for seed white noise power (from top to bottom, $M_H = 10^{-6} M_\odot, \mbox{ } 10^{-3} M_\odot, \mbox{ }  1 M_\odot, \mbox{ }  10^3 M_\odot$).   The $10^{-3}$ and $10^{-6} M_\odot$ curves are indistinguishable from each other and the predictions of CDM (the dashed line in fig.~\ref{spin-temp-cdm}).  For comparison to the background CMB temperature, $T_{CMB}$ is the solid line. The coupling of all kinetic temperatures at $z \gtrsim 200$ to the CMB temperature is the result of coupling of photons and baryons at high redshift.}
\label{early-warm}
\end{figure}

\begin{figure}
\includegraphics[width=8.5cm]{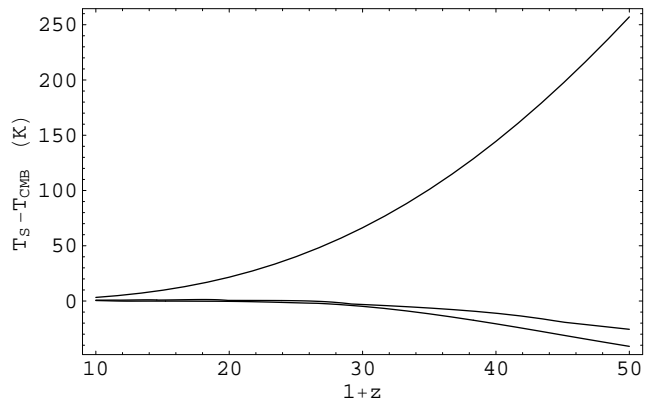}
\caption{Zoom in of fig.~\ref{early-warm} at redshifts more accessible to 21 cm.  The difference between the spin and CMB temperatures are shown for seed white noise power, from bottom to top, $M_H =  10^{-3} M_\odot, \mbox{ }  1 M_\odot, \mbox{ }  10^3 M_\odot$).   The $10^{-3}$ curve is indistinguishable from the predictions of CDM (the short dashed line in fig.~\ref{spin-temp-cdm}).}
\label{early-warm-low}
\end{figure}

The addition of the white noise power to the standard $\Lambda$CDM fluctuations from inflation leads to larger dark matter density fluctuations on small scales.  As a result, dark matter structures form earlier than in standard $\Lambda$CDM.  The virial temperature of early-formed dark matter structures is continually rising with decreasing redshift (see fig.~\ref{white-noise-temp}).  As was shown in fig.~\ref{white-noise-temp}, when that virial temperature rises above the baryon temperature, the baryons clump onto the just-formed dark matter structures.  The virial temperature of these dark matter structures is higher  than in standard $\Lambda$CDM, so that the kinetic temperature of the gas collapsing onto the virialized dark matter is also higher than in standard $\Lambda$CDM.  As a result, the spin temperature of the gas tracks the hotter baryons. Of course, in the limit of sufficiently small white noise power, the spin temperature should reduce to the CDM result.  This situation is shown in fig.~\ref{early-warm}, with a zoom in at lower temperatures in fig.~\ref{early-warm-low}.  
 For sufficiently high $M_H$ the spin temperature may actually be {\em greater} than the CMB photons when it would be less in $\Lambda$CDM.  

Thus for large enough amplitude white noise power the hyperfine transition is seen in emission, rather than absorption (that is, $T_S > T_{CMB}$). For $M_H=10^3 M_\odot$, the gas is {\em always} seen in emission.
Even with (typical models of) heating from star formation included, this behavior {\em never} occurs at   intermediate or high redshifts ($z \gtrsim 15$) with standard CDM. 
In this high power case, the 21cm signals are also larger, by a large factor, than those predicted in standard models, especially at higher redshift. Thus, dark matter models in this range should be significantly constrained by tomographic experiments that reach the $\Lambda$CDM sensitivity level even at $z\approx 15$.

We note that 21cm tomographic  experiments measure differential anisotropy, and are designed to seek particular signatures of structures in frequency and angle. The basic design of these experiments does not   distinguish between emission and absorption in the linear regime. The main signatures of white noise then arise from having  departure of $|T_S - T_\gamma |$ from $\Lambda$CDM, particularly from effects of inhomogeneous (virial or stellar)  heating at  redshifts $z>20$ where $\Lambda$CDM makes clean predictions, because no nonlinear structures or stars have yet formed. In addition,  there are also   experiments under development to test the specific spectral signature of 21 cm absorption and emission; these would be able to detect more directly the radical departure of these models from $\Lambda$CDM predictions for spin temperature.

In the case of small seed structures (corresponding to smaller amplitude white noise), heating by halos alone may not be sufficient to raise the spin temperature above $T_{CMB}$, but, as can be seen from fig.~\ref{early-warm}, the spin temperature is nevertheless higher than standard CDM.    As the radiation from the first stars may serve to further raise the kinetic temperature, and hence the spin temperature, of the gas, the curves in figs.~\ref{early-warm},\ref{early-warm-low} should thus be interpreted as a lower bound on the heating effect. At $z<20$ however, where the first tomographic experiments will operate,  it may be difficult to distinguish these effects from heating by stars.

The white noise hierarchy generates a rising mean temperature but also a range of temperatures.  Not all baryons are heated to the (continually rising) dark matter virial temperature as the hierarchy forms.  Particularly if cooling is inefficient and stars do not form, some baryons  remain in the minihalos in which they first collapsed, which have a lower virial temperature than those currently collapsing and shock heating.  As a result, there is range of temperatures and densities contributing to  21 cm absorption or emission at high redshift: baryons which are heated to the virial temperature of newly formed dark matter halos in the hierarchy, as well as baryons which remain in the cooler minihalos.  While the baryons in the minihalos are cooler, the density of baryons in the minihalos is much higher than in the heated, diffuse gas.  Once they are formed, their kinetic and spin temperature remains low, while the CMB photon temperature continues to drop with redshift.  As a result, these objects will result in less absorption of CMB photons at high redshift as compared to standard CDM, and eventually this weakened absorption will turn into emission, rather than absorption, of the 21 cm line.
We summarize this situation in table~I.

%The discussion here has included only the spin temperature effects from background CMB photons interacting with the primordially neutral hydrogen.  Though we have not considered them here in detail, there will also be  additional effects of the formation of the early stars which heat and partially ionize the gas.  For high mass seeds ($M_h \sim 10^3 M_\odot$), star formation   occurs as early as $z_{eq}$, as can be seen in fig.\ref{white-noise-temp}.  Early star formation and ionization may alter the evolution shown in fig.~\ref{early-warm} through this heating and ionization, though not in the same way as standard CDM.  In standard CDM, the first stars form  and  begin to heat the gas at $z \sim 15-20$, and reionize the gas   at $z \sim 7$.  This heating  eventually raises the spin temperature of the gas above the microwave background, so that 21 cm emission (instead of absorption) is observed, until the gas is totally reionized around $z \sim 6$ so that the 21 cm line disappears. The early star formation occuring in models with added white noise   results in additional early heating.\footnote{Heating of free electrons also produces small scale anisotropy and spectral distortion via the Sunyaev-Zeldovich effect which is in principle observable.}
 
Another  phenomenological feature  in the tomographic signal worth mentioning is the angular white noise from inhomogeneous heating. The main new contribution to the signal we have been discussing assumes that the spatial density  distribution on large scales is the same as standard CDM,  and   the departure from this comes about from the significant change in mean spin temperature of the diffuse HI gas. But an additional signal arises from the fact that the spin temperature itself has a white noise spatial component from the random heating,  that can in principle dominate over the density perturbations on large scales.  This effect adds yet more power to the tomography signal which is not calculated here.

\begin{table}[tbh]
\begin{tabular}{|c|c|c|c|} \hline
$M_H$ ($M_\odot$) & $z_b$ & minihalo $T_S$ (K) & $z_{emit}$ \\ \hline
$10^3$ & 600 & 1500 & 600 \\ \hline
1 & 50 & 90 & 30 \\ \hline
$10^{-3}$ & 15 & 40 & 15 \\ \hline 
$10^{-6}$ & 15 & 40 & 15 \\ \hline
\end{tabular}
\label{minihalo}
\caption{For a given white noise power spectrum associated with mass $M_h$, the redshift at which baryons begin accreting on minihalos, $z_b$; the minihalo spin temperature, $T_S$; and the redshift, $z_{emit}$, at which the baryons accreted onto the minihalo begin emitting rather than absorbing the 21 cm line.}
\end{table}

\section{Conclusion}

We have shown that the addition of white noise to the inflationary power spectrum can alter the history of structure formation and baryon collapse, and lead to modified 21 cm emission and absorption from gas at high redshift.  The effect of the white noise is to seed early bottom-up hierarchical structure formation onto which baryons can collapse, leading to early heating of the baryons through accretion onto   dark matter halos.  This early heating leads to  higher kinetic and spin temperatures, with  departures from standard CDM at redshifts $z \gtrsim 30$ if the seed dark matter halo has mass $\gtrsim 1 M_\odot$.  At higher seed mass $\gtrsim 10 M_\odot$ the additional white noise   also leads to early star formation;  although the observable effects  of that are very model dependent, there are clearly departures expected from $\Lambda$CDM. For $M_H< 1 M_\odot$ the effects appear to be hard to detect since the halo temperature is cooler than the baryon temperature until the   standard inflationary fluctuations begin their collapse.

The observation of such macroscopic astrophysical effects would hint at exotic microphysics in the dark sector.  Standard thermal relics, such as WIMPs and axions, do not give rise to deviations from the inflationary power spectrum that are observable by these techniques.  The most natural source for the extra small scale noise arises from other dark matter dynamics, such as a scalar field phase transition. The signatures discussed here complement effects of  similar minihalo masses that might be observed in gravitational lensing and other lower-redshift effects \cite{Zurek:2006sy}.

\acknowledgements
We are grateful to N. Afshordi, M. Morales and E. Siegel for helpful comments.  This work was supported in part by the U.S. Department of Energy under grant No. DE-FG02-95ER40896.

\bibliography{smallscalepower}

\begin{thebibliography}{24}
\expandafter\ifx\csname natexlab\endcsname\relax\def\natexlab#1{#1}\fi
\expandafter\ifx\csname bibnamefont\endcsname\relax
  \def\bibnamefont#1{#1}\fi
\expandafter\ifx\csname bibfnamefont\endcsname\relax
  \def\bibfnamefont#1{#1}\fi
\expandafter\ifx\csname citenamefont\endcsname\relax
  \def\citenamefont#1{#1}\fi
\expandafter\ifx\csname url\endcsname\relax
  \def\url#1{\texttt{#1}}\fi
\expandafter\ifx\csname urlprefix\endcsname\relax\def\urlprefix{URL }\fi
\providecommand{\bibinfo}[2]{#2}
\providecommand{\eprint}[2][]{\url{#2}}

\bibitem[{\citenamefont{Spergel et~al.}(2006)}]{Spergel:2006hy}
\bibinfo{author}{\bibfnamefont{D.~N.} \bibnamefont{Spergel}}
  \bibnamefont{et~al.} (\bibinfo{year}{2006}), \eprint{astro-ph/0603449}.

\bibitem[{\citenamefont{Adelman-McCarthy
  et~al.}(2006)}]{Adelman-McCarthy:2005se}
\bibinfo{author}{\bibfnamefont{J.~K.} \bibnamefont{Adelman-McCarthy}}
  \bibnamefont{et~al.} (\bibinfo{collaboration}{SDSS}),
  \bibinfo{journal}{Astrophys. J. Suppl.} \textbf{\bibinfo{volume}{162}},
  \bibinfo{pages}{38} (\bibinfo{year}{2006}), \eprint{astro-ph/0507711}.

\bibitem[{\citenamefont{Cole et~al.}(2005)}]{Cole:2005sx}
\bibinfo{author}{\bibfnamefont{S.}~\bibnamefont{Cole}} \bibnamefont{et~al.}
  (\bibinfo{collaboration}{The 2dFGRS}), \bibinfo{journal}{Mon. Not. Roy.
  Astron. Soc.} \textbf{\bibinfo{volume}{362}}, \bibinfo{pages}{505}
  (\bibinfo{year}{2005}), \eprint{astro-ph/0501174}.

\bibitem[{\citenamefont{{McDonald} et~al.}(2006)\citenamefont{{McDonald},
  {Seljak}, {Burles}, {Schlegel}, {Weinberg}, {Cen}, {Shih}, {Schaye},
  {Schneider}, {Bahcall} et~al.}}]{2006ApJS..163...80M}
\bibinfo{author}{\bibfnamefont{P.}~\bibnamefont{{McDonald}}},
  \bibinfo{author}{\bibfnamefont{U.}~\bibnamefont{{Seljak}}},
  \bibinfo{author}{\bibfnamefont{S.}~\bibnamefont{{Burles}}},
  \bibinfo{author}{\bibfnamefont{D.~J.} \bibnamefont{{Schlegel}}},
  \bibinfo{author}{\bibfnamefont{D.~H.} \bibnamefont{{Weinberg}}},
  \bibinfo{author}{\bibfnamefont{R.}~\bibnamefont{{Cen}}},
  \bibinfo{author}{\bibfnamefont{D.}~\bibnamefont{{Shih}}},
  \bibinfo{author}{\bibfnamefont{J.}~\bibnamefont{{Schaye}}},
  \bibinfo{author}{\bibfnamefont{D.~P.} \bibnamefont{{Schneider}}},
  \bibinfo{author}{\bibfnamefont{N.~A.} \bibnamefont{{Bahcall}}},
  \bibnamefont{et~al.}, \bibinfo{journal}{{Astorphys. J. Supp.}}
  \textbf{\bibinfo{volume}{163}}, \bibinfo{pages}{80} (\bibinfo{year}{2006}),
  \eprint{astro-ph/0405013}.

\bibitem[{\citenamefont{Loeb and Zaldarriaga}(2005)}]{Loeb:2005pm}
\bibinfo{author}{\bibfnamefont{A.}~\bibnamefont{Loeb}} \bibnamefont{and}
  \bibinfo{author}{\bibfnamefont{M.}~\bibnamefont{Zaldarriaga}},
  \bibinfo{journal}{Phys. Rev.} \textbf{\bibinfo{volume}{D71}},
  \bibinfo{pages}{103520} (\bibinfo{year}{2005}), \eprint{astro-ph/0504112}.

\bibitem[{\citenamefont{Diemand et~al.}(2005)\citenamefont{Diemand, Moore, and
  Stadel}}]{Diemand:2005vz}
\bibinfo{author}{\bibfnamefont{J.}~\bibnamefont{Diemand}},
  \bibinfo{author}{\bibfnamefont{B.}~\bibnamefont{Moore}}, \bibnamefont{and}
  \bibinfo{author}{\bibfnamefont{J.}~\bibnamefont{Stadel}},
  \bibinfo{journal}{Nature.} \textbf{\bibinfo{volume}{433}},
  \bibinfo{pages}{389} (\bibinfo{year}{2005}), \eprint{astro-ph/0501589}.

\bibitem[{\citenamefont{Kolb and Tkachev}(1996)}]{Kolb:1995bu}
\bibinfo{author}{\bibfnamefont{E.~W.} \bibnamefont{Kolb}} \bibnamefont{and}
  \bibinfo{author}{\bibfnamefont{I.~I.} \bibnamefont{Tkachev}},
  \bibinfo{journal}{Astrophys. J.} \textbf{\bibinfo{volume}{460}},
  \bibinfo{pages}{L25} (\bibinfo{year}{1996}), \eprint{astro-ph/9510043}.

\bibitem[{\citenamefont{Spergel and Steinhardt}(2000)}]{Spergel:1999mh}
\bibinfo{author}{\bibfnamefont{D.~N.} \bibnamefont{Spergel}} \bibnamefont{and}
  \bibinfo{author}{\bibfnamefont{P.~J.} \bibnamefont{Steinhardt}},
  \bibinfo{journal}{Phys. Rev. Lett.} \textbf{\bibinfo{volume}{84}},
  \bibinfo{pages}{3760} (\bibinfo{year}{2000}), \eprint{astro-ph/9909386}.

\bibitem[{\citenamefont{Kaplinghat et~al.}(2000)\citenamefont{Kaplinghat, Knox,
  and Turner}}]{Kaplinghat:2000vt}
\bibinfo{author}{\bibfnamefont{M.}~\bibnamefont{Kaplinghat}},
  \bibinfo{author}{\bibfnamefont{L.}~\bibnamefont{Knox}}, \bibnamefont{and}
  \bibinfo{author}{\bibfnamefont{M.~S.} \bibnamefont{Turner}},
  \bibinfo{journal}{Phys. Rev. Lett.} \textbf{\bibinfo{volume}{85}},
  \bibinfo{pages}{3335} (\bibinfo{year}{2000}), \eprint{astro-ph/0005210}.

\bibitem[{\citenamefont{Feng et~al.}(2003)\citenamefont{Feng, Rajaraman, and
  Takayama}}]{Feng:2003uy}
\bibinfo{author}{\bibfnamefont{J.~L.} \bibnamefont{Feng}},
  \bibinfo{author}{\bibfnamefont{A.}~\bibnamefont{Rajaraman}},
  \bibnamefont{and} \bibinfo{author}{\bibfnamefont{F.}~\bibnamefont{Takayama}},
  \bibinfo{journal}{Phys. Rev.} \textbf{\bibinfo{volume}{D68}},
  \bibinfo{pages}{063504} (\bibinfo{year}{2003}), \eprint{hep-ph/0306024}.

\bibitem[{\citenamefont{Strigari et~al.}(2006)\citenamefont{Strigari,
  Kaplinghat, and Bullock}}]{Strigari:2006jf}
\bibinfo{author}{\bibfnamefont{L.~E.} \bibnamefont{Strigari}},
  \bibinfo{author}{\bibfnamefont{M.}~\bibnamefont{Kaplinghat}},
  \bibnamefont{and} \bibinfo{author}{\bibfnamefont{J.~S.}
  \bibnamefont{Bullock}} (\bibinfo{year}{2006}), \eprint{astro-ph/0606281}.

\bibitem[{\citenamefont{Abazajian et~al.}(2001)\citenamefont{Abazajian, Fuller,
  and Patel}}]{Abazajian:2001nj}
\bibinfo{author}{\bibfnamefont{K.}~\bibnamefont{Abazajian}},
  \bibinfo{author}{\bibfnamefont{G.~M.} \bibnamefont{Fuller}},
  \bibnamefont{and} \bibinfo{author}{\bibfnamefont{M.}~\bibnamefont{Patel}},
  \bibinfo{journal}{Phys. Rev.} \textbf{\bibinfo{volume}{D64}},
  \bibinfo{pages}{023501} (\bibinfo{year}{2001}), \eprint{astro-ph/0101524}.

\bibitem[{\citenamefont{Dodelson and Widrow}(1994)}]{Dodelson:1993je}
\bibinfo{author}{\bibfnamefont{S.}~\bibnamefont{Dodelson}} \bibnamefont{and}
  \bibinfo{author}{\bibfnamefont{L.~M.} \bibnamefont{Widrow}},
  \bibinfo{journal}{Phys. Rev. Lett.} \textbf{\bibinfo{volume}{72}},
  \bibinfo{pages}{17} (\bibinfo{year}{1994}), \eprint{hep-ph/9303287}.

\bibitem[{\citenamefont{Shi and Fuller}(1999)}]{Shi:1998xs}
\bibinfo{author}{\bibfnamefont{X.-D.} \bibnamefont{Shi}} \bibnamefont{and}
  \bibinfo{author}{\bibfnamefont{G.~M.} \bibnamefont{Fuller}},
  \bibinfo{journal}{Phys. Rev.} \textbf{\bibinfo{volume}{D59}},
  \bibinfo{pages}{063006} (\bibinfo{year}{1999}), \eprint{astro-ph/9810075}.

\bibitem[{\citenamefont{Zurek et~al.}(2006)\citenamefont{Zurek, Hogan, and
  Quinn}}]{Zurek:2006sy}
\bibinfo{author}{\bibfnamefont{K.~M.} \bibnamefont{Zurek}},
  \bibinfo{author}{\bibfnamefont{C.~J.} \bibnamefont{Hogan}}, \bibnamefont{and}
  \bibinfo{author}{\bibfnamefont{T.~R.} \bibnamefont{Quinn}}
  (\bibinfo{year}{2006}), \eprint{astro-ph/0607341}.

\bibitem[{\citenamefont{Afshordi et~al.}(2003)\citenamefont{Afshordi, McDonald,
  and Spergel}}]{Afshordi:2003zb}
\bibinfo{author}{\bibfnamefont{N.}~\bibnamefont{Afshordi}},
  \bibinfo{author}{\bibfnamefont{P.}~\bibnamefont{McDonald}}, \bibnamefont{and}
  \bibinfo{author}{\bibfnamefont{D.~N.} \bibnamefont{Spergel}},
  \bibinfo{journal}{Astrophys. J.} \textbf{\bibinfo{volume}{594}},
  \bibinfo{pages}{L71} (\bibinfo{year}{2003}), \eprint{astro-ph/0302035}.

\bibitem[{\citenamefont{Morales}(2005)}]{Morales:2004ca}
\bibinfo{author}{\bibfnamefont{M.~F.} \bibnamefont{Morales}},
  \bibinfo{journal}{Astrophys. J.} \textbf{\bibinfo{volume}{619}},
  \bibinfo{pages}{678} (\bibinfo{year}{2005}), \eprint{astro-ph/0406662}.

\bibitem[{\citenamefont{Morales et~al.}(2006)\citenamefont{Morales, Bowman, and
  Hewitt}}]{Morales:2005qk}
\bibinfo{author}{\bibfnamefont{M.~F.} \bibnamefont{Morales}},
  \bibinfo{author}{\bibfnamefont{J.~D.} \bibnamefont{Bowman}},
  \bibnamefont{and} \bibinfo{author}{\bibfnamefont{J.~N.}
  \bibnamefont{Hewitt}}, \bibinfo{journal}{Astrophys. J.}
  \textbf{\bibinfo{volume}{648}}, \bibinfo{pages}{767} (\bibinfo{year}{2006}),
  \eprint{astro-ph/0510027}.

\bibitem[{\citenamefont{Barkana and Loeb}(2001)}]{Barkana:2000fd}
\bibinfo{author}{\bibfnamefont{R.}~\bibnamefont{Barkana}} \bibnamefont{and}
  \bibinfo{author}{\bibfnamefont{A.}~\bibnamefont{Loeb}},
  \bibinfo{journal}{Phys. Rept.} \textbf{\bibinfo{volume}{349}},
  \bibinfo{pages}{125} (\bibinfo{year}{2001}), \eprint{astro-ph/0010468}.

\bibitem[{\citenamefont{Furlanetto et~al.}(2006)\citenamefont{Furlanetto, Oh,
  and Briggs}}]{Furlanetto:2006jb}
\bibinfo{author}{\bibfnamefont{S.}~\bibnamefont{Furlanetto}},
  \bibinfo{author}{\bibfnamefont{S.~P.} \bibnamefont{Oh}}, \bibnamefont{and}
  \bibinfo{author}{\bibfnamefont{F.}~\bibnamefont{Briggs}},
  \bibinfo{journal}{Phys. Rept.} \textbf{\bibinfo{volume}{433}},
  \bibinfo{pages}{181} (\bibinfo{year}{2006}), \eprint{astro-ph/0608032}.

\bibitem[{\citenamefont{Fan et~al.}(2006)\citenamefont{Fan, Carilli, and
  Keating}}]{Fan:2006dp}
\bibinfo{author}{\bibfnamefont{X.-H.} \bibnamefont{Fan}},
  \bibinfo{author}{\bibfnamefont{C.~L.} \bibnamefont{Carilli}},
  \bibnamefont{and} \bibinfo{author}{\bibfnamefont{B.}~\bibnamefont{Keating}},
  \bibinfo{journal}{Ann. Rev. Astron. Astrophys.}
  \textbf{\bibinfo{volume}{44}}, \bibinfo{pages}{415} (\bibinfo{year}{2006}),
  \eprint{astro-ph/0602375}.

\bibitem[{\citenamefont{Hogan and Rees}(1988)}]{Hogan:1988mp}
\bibinfo{author}{\bibfnamefont{C.~J.} \bibnamefont{Hogan}} \bibnamefont{and}
  \bibinfo{author}{\bibfnamefont{M.~J.} \bibnamefont{Rees}},
  \bibinfo{journal}{Phys. Lett.} \textbf{\bibinfo{volume}{B205}},
  \bibinfo{pages}{228} (\bibinfo{year}{1988}).

\bibitem[{\citenamefont{Kolb and Tkachev}(1993)}]{Kolb:1993zz}
\bibinfo{author}{\bibfnamefont{E.~W.} \bibnamefont{Kolb}} \bibnamefont{and}
  \bibinfo{author}{\bibfnamefont{I.~I.} \bibnamefont{Tkachev}},
  \bibinfo{journal}{Phys. Rev. Lett.} \textbf{\bibinfo{volume}{71}},
  \bibinfo{pages}{3051} (\bibinfo{year}{1993}), \eprint{hep-ph/9303313}.

\bibitem[{\citenamefont{Loeb and Zaldarriaga}(2004)}]{Loeb:2003ya}
\bibinfo{author}{\bibfnamefont{A.}~\bibnamefont{Loeb}} \bibnamefont{and}
  \bibinfo{author}{\bibfnamefont{M.}~\bibnamefont{Zaldarriaga}},
  \bibinfo{journal}{Phys. Rev. Lett.} \textbf{\bibinfo{volume}{92}},
  \bibinfo{pages}{211301} (\bibinfo{year}{2004}), \eprint{astro-ph/0312134}.

\end{thebibliography}


\begin{thebibliography}{11}
\expandafter\ifx\csname natexlab\endcsname\relax\def\natexlab#1{#1}\fi
\expandafter\ifx\csname bibnamefont\endcsname\relax
  \def\bibnamefont#1{#1}\fi
\expandafter\ifx\csname bibfnamefont\endcsname\relax
  \def\bibfnamefont#1{#1}\fi
\expandafter\ifx\csname citenamefont\endcsname\relax
  \def\citenamefont#1{#1}\fi
\expandafter\ifx\csname url\endcsname\relax
  \def\url#1{\texttt{#1}}\fi
\expandafter\ifx\csname urlprefix\endcsname\relax\def\urlprefix{URL }\fi
\providecommand{\bibinfo}[2]{#2}
\providecommand{\eprint}[2][]{\url{#2}}

\bibitem[{\citenamefont{Adelman-McCarthy
  et~al.}(2006)}]{Adelman-McCarthy:2005se}
\bibinfo{author}{\bibfnamefont{J.~K.} \bibnamefont{Adelman-McCarthy}}
  \bibnamefont{et~al.} (\bibinfo{collaboration}{SDSS}),
  \bibinfo{journal}{Astrophys. J. Suppl.} \textbf{\bibinfo{volume}{162}},
  \bibinfo{pages}{38} (\bibinfo{year}{2006}), \eprint{astro-ph/0507711}.

\bibitem[{\citenamefont{Cole et~al.}(2005)}]{Cole:2005sx}
\bibinfo{author}{\bibfnamefont{S.}~\bibnamefont{Cole}} \bibnamefont{et~al.}
  (\bibinfo{collaboration}{The 2dFGRS}), \bibinfo{journal}{Mon. Not. Roy.
  Astron. Soc.} \textbf{\bibinfo{volume}{362}}, \bibinfo{pages}{505}
  (\bibinfo{year}{2005}), \eprint{astro-ph/0501174}.

\bibitem[{\citenamefont{{McDonald} et~al.}(2006)\citenamefont{{McDonald},
  {Seljak}, {Burles}, {Schlegel}, {Weinberg}, {Cen}, {Shih}, {Schaye},
  {Schneider}, {Bahcall} et~al.}}]{2006ApJS..163...80M}
\bibinfo{author}{\bibfnamefont{P.}~\bibnamefont{{McDonald}}},
  \bibinfo{author}{\bibfnamefont{U.}~\bibnamefont{{Seljak}}},
  \bibinfo{author}{\bibfnamefont{S.}~\bibnamefont{{Burles}}},
  \bibinfo{author}{\bibfnamefont{D.~J.} \bibnamefont{{Schlegel}}},
  \bibinfo{author}{\bibfnamefont{D.~H.} \bibnamefont{{Weinberg}}},
  \bibinfo{author}{\bibfnamefont{R.}~\bibnamefont{{Cen}}},
  \bibinfo{author}{\bibfnamefont{D.}~\bibnamefont{{Shih}}},
  \bibinfo{author}{\bibfnamefont{J.}~\bibnamefont{{Schaye}}},
  \bibinfo{author}{\bibfnamefont{D.~P.} \bibnamefont{{Schneider}}},
  \bibinfo{author}{\bibfnamefont{N.~A.} \bibnamefont{{Bahcall}}},
  \bibnamefont{et~al.}, \bibinfo{journal}{{Astorphys. J. Supp.}}
  \textbf{\bibinfo{volume}{163}}, \bibinfo{pages}{80} (\bibinfo{year}{2006}),
  \eprint{astro-ph/0405013}.

\bibitem[{\citenamefont{Spergel et~al.}(2006)}]{Spergel:2006hy}
\bibinfo{author}{\bibfnamefont{D.~N.} \bibnamefont{Spergel}}
  \bibnamefont{et~al.} (\bibinfo{year}{2006}), \eprint{astro-ph/0603449}.

\bibitem[{\citenamefont{Zurek et~al.}(2006)\citenamefont{Zurek, Hogan, and
  Quinn}}]{Zurek:2006sy}
\bibinfo{author}{\bibfnamefont{K.~M.} \bibnamefont{Zurek}},
  \bibinfo{author}{\bibfnamefont{C.~J.} \bibnamefont{Hogan}}, \bibnamefont{and}
  \bibinfo{author}{\bibfnamefont{T.~R.} \bibnamefont{Quinn}}
  (\bibinfo{year}{2006}), \eprint{astro-ph/0607341}.

\bibitem[{\citenamefont{Das and Weiner}(2006)}]{Das:2006ht}
\bibinfo{author}{\bibfnamefont{S.}~\bibnamefont{Das}} \bibnamefont{and}
  \bibinfo{author}{\bibfnamefont{N.}~\bibnamefont{Weiner}}
  (\bibinfo{year}{2006}), \eprint{astro-ph/0611353}.

\bibitem[{\citenamefont{Furlanetto et~al.}(2004)\citenamefont{Furlanetto,
  Zaldarriaga, and Hernquist}}]{Furlanetto:2004ha}
\bibinfo{author}{\bibfnamefont{S.}~\bibnamefont{Furlanetto}},
  \bibinfo{author}{\bibfnamefont{M.}~\bibnamefont{Zaldarriaga}},
  \bibnamefont{and}
  \bibinfo{author}{\bibfnamefont{L.}~\bibnamefont{Hernquist}},
  \bibinfo{journal}{Astrophys. J.} \textbf{\bibinfo{volume}{613}},
  \bibinfo{pages}{16} (\bibinfo{year}{2004}), \eprint{astro-ph/0404112}.

\bibitem[{\citenamefont{Diemand et~al.}(2005)\citenamefont{Diemand, Moore, and
  Stadel}}]{Diemand:2005vz}
\bibinfo{author}{\bibfnamefont{J.}~\bibnamefont{Diemand}},
  \bibinfo{author}{\bibfnamefont{B.}~\bibnamefont{Moore}}, \bibnamefont{and}
  \bibinfo{author}{\bibfnamefont{J.}~\bibnamefont{Stadel}},
  \bibinfo{journal}{Nature.} \textbf{\bibinfo{volume}{433}},
  \bibinfo{pages}{389} (\bibinfo{year}{2005}), \eprint{astro-ph/0501589}.

\bibitem[{\citenamefont{Bertschinger}(2006)}]{Bertschinger:2006nq}
\bibinfo{author}{\bibfnamefont{E.}~\bibnamefont{Bertschinger}},
  \bibinfo{journal}{Phys. Rev.} \textbf{\bibinfo{volume}{D74}},
  \bibinfo{pages}{063509} (\bibinfo{year}{2006}), \eprint{astro-ph/0607319}.

\bibitem[{\citenamefont{Hogan and Rees}(1988)}]{Hogan:1988mp}
\bibinfo{author}{\bibfnamefont{C.~J.} \bibnamefont{Hogan}} \bibnamefont{and}
  \bibinfo{author}{\bibfnamefont{M.~J.} \bibnamefont{Rees}},
  \bibinfo{journal}{Phys. Lett.} \textbf{\bibinfo{volume}{B205}},
  \bibinfo{pages}{228} (\bibinfo{year}{1988}).

\bibitem[{\citenamefont{Kolb and Tkachev}(1993)}]{Kolb:1993zz}
\bibinfo{author}{\bibfnamefont{E.~W.} \bibnamefont{Kolb}} \bibnamefont{and}
  \bibinfo{author}{\bibfnamefont{I.~I.} \bibnamefont{Tkachev}},
  \bibinfo{journal}{Phys. Rev. Lett.} \textbf{\bibinfo{volume}{71}},
  \bibinfo{pages}{3051} (\bibinfo{year}{1993}), \eprint{hep-ph/9303313}.

\end{thebibliography}

\end{document}